\documentclass[
    ,final            
  ]
  {aipproc}

\layoutstyle{6x9}

\begin{document}

\title{QCD dynamics in mesons at soft and hard scales}


\author{T. Nguyen}{
  address={Center for Nuclear Research, Department of Physics,  Kent State 
University, Kent OH 44242, USA}
}

\author{N. A. Souchlas}{
  address={Center for Nuclear Research, Department of Physics,  Kent State 
University, Kent OH 44242, USA}
}

\author{P. C. Tandy}{
  address={Center for Nuclear Research, Department of Physics,  Kent State 
University, Kent OH 44242, USA}
}

\begin{abstract}
Using a ladder-rainbow kernel previously established for the soft scale of light quark hadrons, we explore, within a Dyson-Schwinger approach,  phenomena that mix soft and hard scales of QCD.     The difference between vector and axial vector current correlators is examined to estimate the four quark chiral condensate and the leading distance scale for the onset of non-perturbative phenomena in QCD.    The valence quark distributions, in the pion and kaon,  defined in deep inelastic scattering, and measured in the Drell Yan process, are investigated with the same ladder-rainbow truncation of the Dyson-Schwinger and Bethe-Salpeter equations.
\end{abstract}

\maketitle



A great deal of progress in the QCD modeling of hadron physics has been 
achieved through the use of the ladder-rainbow truncation of the Dyson-Schwinger
equations (DSEs).   The DSEs are the equations of motion of a
quantum field theory.  They form an infinite hierarchy of coupled
integral equations for the Green's functions ($n$-point functions) of
the theory.  Bound states (mesons, baryons) appear as poles in the appropriate
Green's functions, and, e.g., the Bethe-Salpeter bound state equation appears after taking residues in the DSE for the appropriate color singlet vertex. 
For recent reviews on the DSEs and their use in hadron physics, see
Refs.~\cite{Roberts:1994dr,Tandy:1997qf,Alkofer:2000wg,Maris:2003vk}.   

In the Euclidean metric that we use throughout, the DSE for the dressed quark propagator is
\begin{eqnarray}
S(p)^{-1}  &=& Z_2 \, i\,/\!\!\!p + Z_4 \, m(\mu) + Z_1 \int^\Lambda_q \! g^2D_{\mu\nu}(p-q) \, 
        \frac{\lambda^i}{2}\gamma_\mu \, S(q) \, \Gamma^i_\nu(q,p)~,
\label{quarkdse}
\end{eqnarray}
where $D_{\mu\nu}(k)$ is the renormalized dressed-gluon propagator,
$\Gamma^i_\nu(q,p)$ is the renormalized dressed quark-gluon vertex.
We use $\int_q^\Lambda$ to denote $\int^\Lambda  d^4q/(2\pi)^4$ with $\Lambda$ being the
mass scale for translationally invariant regularization.    The renormalization condition is 
$S(p)^{-1}=i\gamma\cdot p+m(\mu)$ at a sufficiently large spacelike
$\mu^2$, with $m(\mu)$ the renormalized mass at renormalization scale $\mu$.
We use \mbox{$\mu=19\,{\rm GeV}$}.   The $Z_i(\mu, \Lambda)$ are renormalization constants.  
Bound state pole residues of the inhomogeneous Bethe-Salpeter equation (BSE) for the relevant vertex, yield the homogeneous BSE bound state equation 
\begin{eqnarray}
\Gamma^{a\bar{b}}(p_+,p_-) &=& \int^\Lambda_q \! K(p,q;P)S^a(q_+)
                                          \Gamma^{a\bar{b}}(q_+,q_-)S^b(q_-)~~,
\label{bse}
\end{eqnarray}
where $K$ is the renormalized $q\bar{q}$ scattering kernel that is
irreducible with respect to a pair of $q\bar{q}$ lines.  Quark
momenta are \mbox{$q_+ =$} \mbox{$q+\eta P$} and \mbox{$q_- =$} \mbox{$q -$}
\mbox{$(1-\eta) P$} where the choice of $\eta$ is equivalent to a definition of relative
momentum $q$; observables should not depend on $\eta$.  The meson momentum 
satisfies $P^2 = -M^2$.



A viable truncation of the infinite set of DSEs should respect
relevant (global) symmetries of QCD such as chiral symmetry, Lorentz
invariance, and renormalization group invariance.  For electromagnetic
interactions and Goldstone bosons we also need to respect color singlet  vector 
and axial vector current conservation.  The rainbow-ladder (LR) truncation achieves these ends by
the replacement
\mbox{$K(p,q;P)  \to -4\pi\,\alpha_{\rm eff}(k^2)\, D_{\mu\nu}^{\rm free}(k)
\textstyle{\frac{\lambda^i}{2}}\gamma_\mu \otimes \textstyle{\frac{\lambda^i}{2}}\gamma_\nu $}
along with the replacement of the DSE kernel  by
\mbox{$ Z_1 g^2 D_{\mu \nu}(k) \Gamma^i_\nu(q,p) \to 
 4\pi\,\alpha_{\rm eff}(k^2) \, D_{\mu\nu}^{\rm free}(k)\, \gamma_\nu
                                        \textstyle\frac{\lambda^i}{2} $}
where $k=p-q$, and $\alpha_{\rm eff}(k^2)$ is an effective running
coupling.   This truncation is the first term in a systematic
expansion~\cite{Bender:1996bb,Bhagwat:2004hn} of $K$; asymptotically, it reduces to leading-order perturbation theory.
These two truncations are mutually consistent: together they produce color singlet vector and axial-vector vertices satisfying their respective Ward identities.  This
ensures that the chiral limit ground state pseudoscalar bound states
are the massless Goldstone bosons from chiral symmetry
breaking~\cite{Maris:1998hd,Maris:1997tm}; and
ensures electromagnetic current conservation~\cite{Roberts:1996hh}.  

\begin{table}
\caption{DSE results~\protect\cite{Maris:1999nt} for pseudoscalar and vector meson masses and electroweak decay constants, together with experimental data~\protect\cite{PDG04}.   Units are GeV except where indicated.   Quantities marked by $\dagger$ are fitted with the indicated current quark masses and the infrared strength parameter of the ladder-rainbow kernel.  \label{Table:model} }

\begin{tabular}{|l|cc|cc|cc|cc|cc|} \hline 
  \multicolumn{2}{|c|}{ }   & \multicolumn{3}{c|}{$m^{u=d}_{\mu=1 {\rm GeV}}$}  &  \multicolumn{3}{c|}{$m^{s}_{\mu=1 {\rm GeV}}$}  &   \multicolumn{3}{c|}{- $\langle \bar q q \rangle^0_{\mu=1 {\rm GeV}}$}    \\   \hline
 \multicolumn{2}{|c|}{ expt }   &  \multicolumn{3}{c|}{ 3 - 6 MeV}   &  \multicolumn{3}{c|}{  80 - 130 MeV } &  \multicolumn{3}{c|}{ (0.24 GeV)$^3$ }       \\     
\multicolumn{2}{|c|}{  calc }  &   \multicolumn{3}{c|}{ 5.5 MeV}  &  \multicolumn{3}{c|}{  125 MeV }  &   \multicolumn{3}{c|}{  (0.241 GeV)$^{3\dagger}$  }       \\ \hline
        & $m_\pi$ & $f_\pi$ & $m_K$ & $f_K$   &   $m_\rho$ &  $f_\rho$  & $m_K^\star$ & $f_K^\star$ &  $m_\phi$ &  $f_\phi$
 \\ \hline 
 expt  &   0.138  &  0.131  &   0.496   &   0.160 &   0.770  &  0.216  &   0.892 &  0.225   &   1.020   &  0.236    \\  
 calc  &   0.138$^\dagger$ &  0.131$^\dagger$ & 0.497$^\dagger$ &  0.155  &  0.742 & 0.207 & 0.936 & 0.241  &  1.072     &    0.259   \\ \hline  
\end{tabular}
\end{table}

We employ the ladder-rainbow kernel found to be successful in earlier 
work for light quarks~\cite{Maris:1997tm,Maris:1999nt}.   It can be written
\mbox{$\alpha_{\rm eff}(k^2) =  \alpha^{\rm IR}(k^2) + \alpha^{\rm UV}(k^2) $}.
The IR term implements the strong infrared enhancement in the region
\mbox{$0 < k^2 < 1\,{\rm GeV}^2$} required for sufficient dynamical
chiral symmetry breaking.   The UV term preserves the one-loop renormalization group behavior of QCD: \mbox{$\alpha_{\rm eff}(k^2) \to \alpha_s(k^2)^{\rm 1 loop}(k^2)$} in the ultraviolet with 
\mbox{$N_f=4$} and \mbox{$\Lambda_{\rm QCD} = 0.234\,{\rm GeV}$}.   The strength of $\alpha^{\rm IR}$ along with two quark masses are fitted to $\langle\bar q q\rangle $, $m_{\pi/K}$.   Selected light quark meson results are displayed in Table~\ref{Table:model}.   The infrared kernel component is phenomenological because QCD is unsolved in such a non-perturbative domain. To help replace such phenomenology by specific mechanisms, it is necessary to first  characterize its performance in new domains.


Quark helicity and chirality in QCD are increasingly good quantum numbers at short distances or at momentum scales significantly larger than any mass scale.   One manifestation of this is that, for chiral quarks, the correlator of a pair of vector currents is identical to the corresponding correlator of a pair of axial vector currents to all finite orders of pQCD.   Non-perturbatively,  the difference of such correlators measures chirality flips, and the leading non-zero ultraviolet contribution identifies the leading non-perturbative phenomenon in QCD.     This is the four quark 
condensate~\cite{Narison:1989aq}.    Here, the vector correlator is formulated as the loop integral
\begin{equation}
\Pi_{\mu\nu}^{V}(P) = \int d^{4}x \; {\rm e}^{iP \cdot x} 
\langle 0|T \, j_\mu(x)\, j^+_\nu(0)|0 \rangle = -\int^\Lambda_q 
Tr\lbrace \gamma_\mu S(q_+) \Gamma^{V}_{\nu} (q,P)S(q_-)\rbrace ~~,
\label{Eq:vcorr}
\end{equation}
where $\Lambda $ indicates regularization, e.g., by the Pauli-Villars method, and $\Gamma^{V}_{\nu}$ is the dressed vector vertex.   The axial vector correlator is formulated in an analogous way and we directly calculate the difference correlator which does not require ultraviolet regularization.   
We consider the transverse difference \mbox{$\Pi_{T}^{V-A}(P^{2}) = \Pi_{T}^{V}(P^{2}) - \Pi_{T}^{A}(P^{2}) $}.  

The leading non-perturbative contribution  to $\Pi_{T}^{V-A}$ starts with 
dimension $d =6$ and involves the four-quark condensate in the form~\cite{Dominguez:1998wy,Dominguez:2003dr}
\begin{equation}
\label{eq:c4cm}
\Pi_{T}^{V-A}(P^{2}) = -\frac{32 \pi}{9} 
\frac{\alpha_{s} \langle \bar{q}q\bar{q}q\rangle }{P^{6}}
\{1+\frac{\alpha_{s}(P^{2})}{4\pi}[\frac{247}{12}+ {\rm ln} (\frac{\mu^{2}}{P^{2}})]
\} + O(\frac{1}{P^{8}})~~.
\end{equation}
Our numerical calculation of $P^6\, \Pi_{T}^{V-A}(P^{2})$ identifies a leading ultraviolet constant reasonably well.   The four quark condensate 
$\langle \bar{q}q\bar{q}q\rangle $ extracted via Eq.~(\ref{eq:c4cm}) is 65\% greater than the common vacuum saturation assumption $\langle \bar{q}q\rangle^2$ at the renormalization scale \mbox{$\mu = 19$}~GeV used in this work.  
The low $P^2$ limit provides a reasonable account of the first Weinberg sum rule~\cite{Weinberg:1967kj, Dorokhov:2003kf}:   
\mbox{$P^{2} \,\Pi_{T}^{V-A}(P^{2})|_{P^{2} \to 0}  = - f_{\pi}^{2} $}, in the \mbox{$f_\pi = 0.0924$} GeV convention.    This limit is due to $\Pi_{T}^{A}$ only and  we obtain  \mbox{$f_\pi = 0.09$} GeV that way.   Our results are consistent with the  second Weinberg sum rule~\cite{Weinberg:1967kj} \mbox{$P^4 \,\Pi_{T}^{V-A}(P^{2})|_{P^{2} \to \infty}  =0$}.    The Das-Guralnik-Mathur-Low-Young sum rule~\cite{Das:1967it} relates 
\mbox{$\int_{0}^{\infty}{dP^{2}}\, P^{2}\,\Pi_{T}^{V-A}(P^{2})$} to  the strong component of $m_{\pi^{\pm}} - m_{\pi^{0}}$.   We obtain 4.86 MeV for this mass difference in comparison with $4.43 \pm 0.03$  from experiment.   


Data for the momentum-fraction probability distributions of quarks and gluons in the pion have primarily been inferred from Drell-Yan processes in pion-nucleon collisions~\cite{Badier:1983mj,Betev:1985pg,Conway:1989fs}.    For a recent review of nucleon and pion parton distributions see Ref.~\cite{Holt:2010vj}.
\begin{figure}[th] 
\hspace*{4mm}
\includegraphics[height=0.26\textheight]{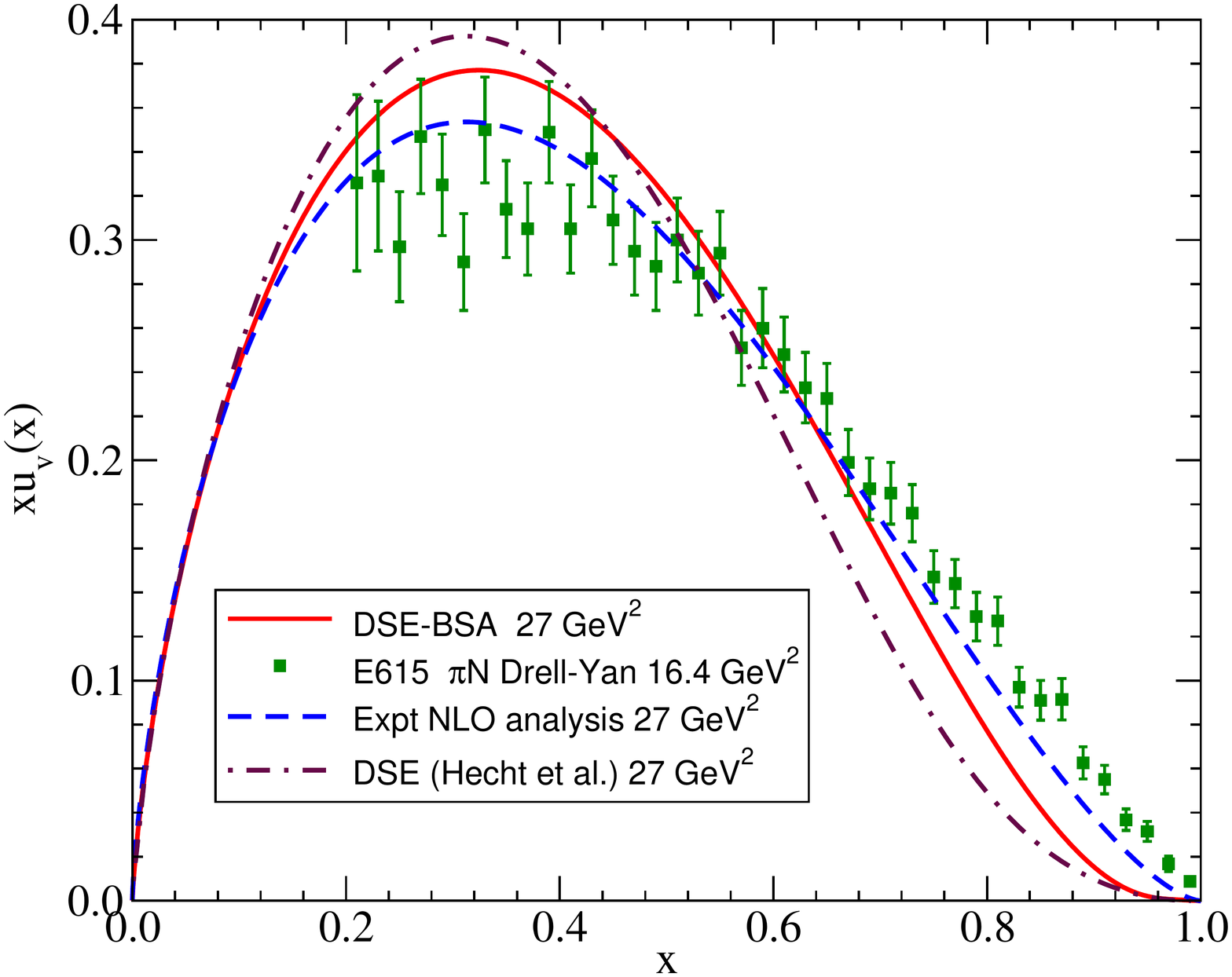}\hspace*{2mm}
\includegraphics[height=0.28\textheight]{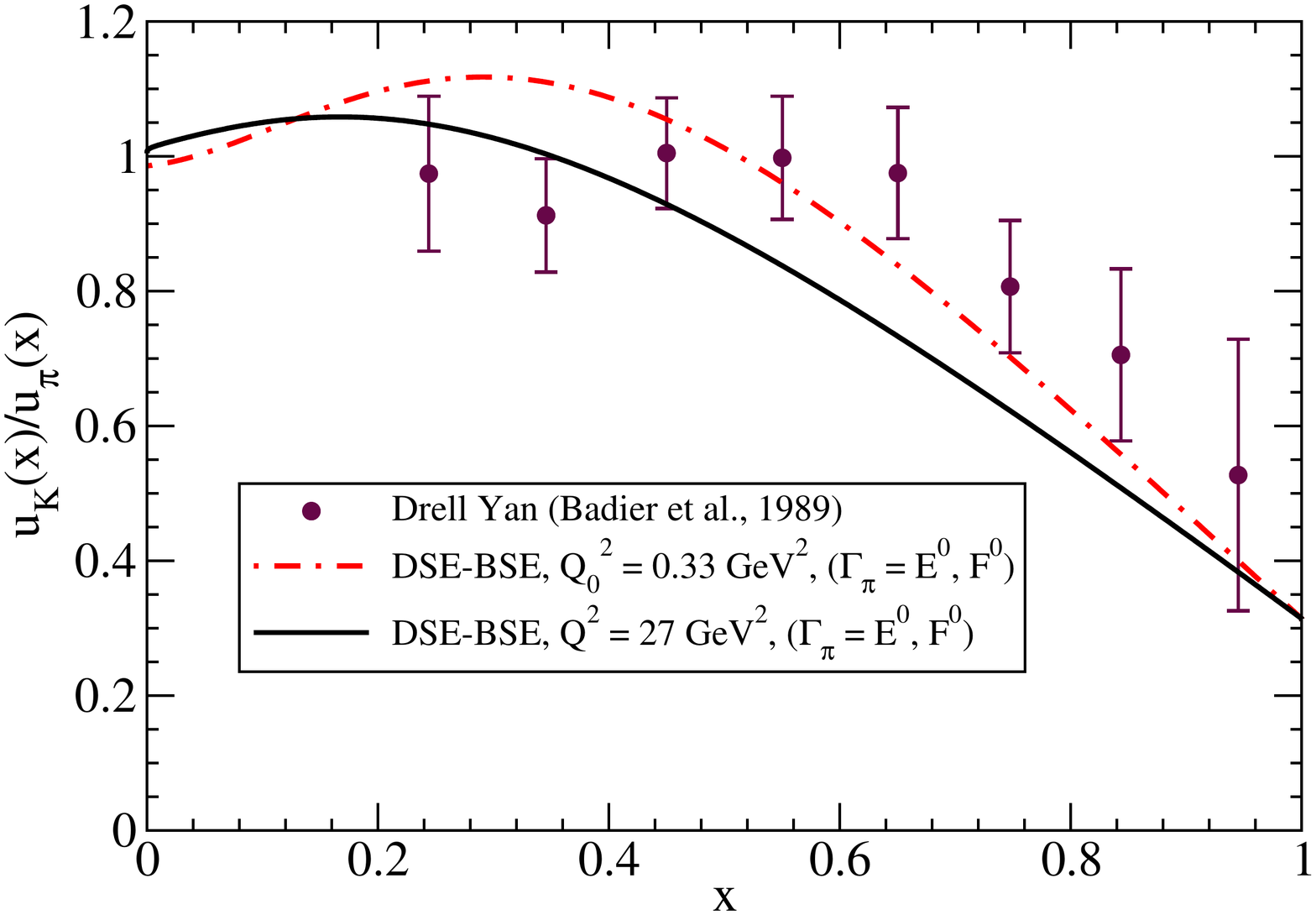}

\caption{{\it Left Panel}:    Pion valence quark distribution evolved to (5.2~GeV)$^2$.  Solid line is the full DSE-BSA calculation~\protect\cite{Nguyen_PhD09}; dot-dashed line is the semi-phenomenological DSE-based calculation of Hecht et al.~\protect\cite{Hecht:2000xa};   experimental data points are from~\protect\cite{Conway:1989fs} at scale (4.05~GeV)$^2$; the dashed line is the recent NLO re-analysis of the experimental data~\protect\cite{Wijesooriya:2005ir}.   {\it Right Panel}:   The ratio of u-quark distributions in the kaon and pion.   The solid line is our preliminary result from DSE-BSE 
calculations~\protect\cite{Nguyen:inprep10,Nguyen_PhD09,Holt:2010vj};  the experimental data is from~\protect\cite{Badier:1980jq,Badier:1983mj}.  \label{fig:pi_DSE+ratio} }
\end{figure} 
Lattice-QCD is restricted to  low moments of the distributions, not the
distributions themselves~\cite{Best:1997qp}.    Model calculations of deep inelastic scattering 
(DIS) parton distribution functions are challenging because it is necessary to have hard scale perturbative QCD features coexisting with a covariant nonperturbative model of soft scale bound states.     In DIS calculations within the Nambu--Jona-Lasinio model~\cite{Shigetani:1993dx,Weigel:1999pc,Bentz:1999gx}, difficulties include a point structure for the pion and a marked sensitivity to the regularization procedure.     In constituent quark models~\cite{Szczepaniak:1993uq,Frederico:1994dx}, and  instanton-liquid models~\cite{Dorokhov:2000gu}, it is difficult to have pQCD elements join smoothly with nonperturbative aspects.   An approach styled after the successful DSE treatment of the 
pion electromagnetic form factor~\cite{Maris:2000sk,Maris:1999bh} can overcome these difficulties.  

The Bjorken kinematic limit of DIS selects the most singular behavior of a correlator of quark fields of the target with light-like separation $z^2 \sim 0 $.   With incident photon momentum along the negative 
3-axis, the kinematics selects \mbox{$z^+ \sim z_\perp \sim 0$}  leaving  $z^-$ as the finite distance conjugate to quark momentum component $xP^+$, where  \mbox{$ x = Q^2/2P \cdot q$} is the Bjorken variable, \mbox{$q^2 = -Q^2$} is the spacelike virtuality of the photon, and $P$ is the target momentum. To leading  order in the operator product expansion, the associated probability amplitude $q_f(x)$, characteristic of the target,  is  given by the  correlator~\cite{Jaffe:1985je,Jaffe:1983hp} 
\begin{equation}
q_f(x) = \frac{1}{4 \pi} \int  d z^- e^{i x P^+ z^- } 
\langle \pi(P) | \bar{\psi}_f (z^-) \gamma^+ \psi_f(0) | \pi(P) \rangle ~~~, 
\label{Mink_dis_x}
\end{equation}
where $f$ is a flavor label.
Here the 4-vector components that arise naturally are \mbox{$a^\pm = (a^0 \pm a^3)/\sqrt{2}$}.      This  probability amplitude is invariant, it  can easily be given a manifestly covariant formulation, and its interpretation is perhaps simplest in the infinite momentum frame where $q_f(x)$  is the quantum mechanical probability that a single parton  has momentum fraction 
$x$~\cite{Ellis:1991qj}.   Note that \mbox{$q_f(x) = - q_{\bar{f}}(-x) $}, and that the valence quark amplitude is \mbox{$ q^v_f(x) = q_f(x) - q_{\bar{f}}(x)$}.   It follows from Eq.~(\ref{Mink_dis_x}) that 
\mbox{$ \int_0^1 dx \,  q^v_f(x) = $} \mbox{$ \langle \pi(P) | J^+_f(0) | \pi(P) \rangle /2P^+ = F_\pi(0) = 1$}.   Approximate treatments should at least preserve  vector current conservation to automatically obtain the correct normalization for valence quark number.

In a momentum representation, $q_f(x)$ can be written as the evaluation of a special Feynman diagram~\cite{Jaffe:1985je,Mineo:1999eq}
\mbox{$q_f(x) = \frac{1}{2} \int_k^\Lambda \delta(k^+ - xP^+) \; {\rm tr}_{\rm cd} [\gamma^+ \, G(k,P)]$}
where $G(k,P)$ represents the forward $\bar q$-target scattering amplitude.   Ladder-rainbow truncation, which selects the valence $q \bar q$ structure of the pion, yields  
\begin{equation}
q^v_f(x) = \frac{i}{2} \int_p^\Lambda {\rm tr}_{\rm cd} [ \Gamma_\pi(p,P) \, S(p)\, \Gamma^+(p;x) \, S(p) \,\Gamma_\pi(p,P)\, S(p-P) ]~~~,  
\label{Mink_dis_LR_Ward}
\end{equation}
where ${\rm tr}_{\rm cd}$ denotes a color and Dirac trace, and $\Gamma^+(p;x)$ is a generalization of the dressed vertex that a zero momentum photon has with a quark.  It satisfies the usual  inhomogeneous BSE integral equation  (here with a LR kernel) except that the inhomogeneous term is 
$\gamma^+ \, \delta(p^+ - xP^+)$.   This selection of LR dynamics exactly parallels the symmetry-preserving dynamics of the corresponding treatment of the pion charge form factor at \mbox{$q^2 = 0 $} wherein the vector current is conserved by use of ladder dynamics at all three vertices and rainbow dynamics for all 3 quark propagators~\cite{Maris:2000sk,Maris:1999bh}.   Here the number of valence $u$-quarks (or $\bar d$) in the pion is automatically unity since the structure of Eq.~(\ref{Mink_dis_LR_Ward}), along with the canonical  normalization of the $q \bar q$ BS amplitude $\Gamma_\pi(p,P)$,  ensures \mbox{$ \int_0^1 dx \,  q^v_u(x) = 1$} because $ \int_0^1 dx \, \Gamma^+(p;x)$ gives the Ward Identity vertex.  

Eq.~(\ref{Mink_dis_LR_Ward}) is in Minkowski metric so as to satisfy the constraint on $p^+$, but LR dynamical information on the various non-perturbative elements such as $S(p)$ and  $\Gamma_\pi(p,P)$ is available only in Euclidean metric~\cite{Maris:1999nt}.   Since $q_f(x)$ is obtained from the hadron tensor $W^{\mu \nu}$ which in turn can be formulated from the discontinuity  \mbox{$T^{\mu\nu}(\epsilon) -  T^{\mu\nu}(-\epsilon)$}, we observe that all enclosed singularities from the difference of Wick rotations cancel except for the cut that defines the object of interest.    With use of numerical solutions for dressed propagators  and BS amplitudes, that give an accurate account of light quark hadrons, our DIS calculations  significantly extend  the exploratory study made in Ref.~\cite{Hecht:2000xa}.   That work employed phenomenological parameterizations of these elements.  

In Fig.~\ref{fig:pi_DSE+ratio} we display our DSE result for the valence $u$-quark distribution evolved to $Q^2 = (5.2~{\rm GeV})^2$ in comparison with $\pi N$ Drell-Yan data~\cite{Conway:1989fs} with a scale quoted as  $Q^2 > (4.05~{\rm GeV})^2$.   We also compare with  a recent NLO reanalysis of the data at scale $Q^2 = (5.2~{\rm GeV})^2$.   The distribution at the model scale $Q_0^2$ is evolved higher by leading order DGLAP.    The model scale is found to be \mbox{$Q_0 = 0.57 $}~GeV by matching the $x^n$ moments for $n=1,2,3$ to the experimental values given independently at (2~GeV)$^2$~\cite{Sutton:1991ay}.   
Our momentum sum rule result \mbox{$\int_0^1 dx x(u_\pi + \bar{d}_\pi ) = $} \mbox{$0.74 $} at $Q_0$ clearly show that in a covariant approach the retardation effects of one gluon exchange
assign some of the momentum to gluons.   The corresponding momentum sum for the kaon is $0.76$.

The ratio $u_K/u_\pi$ measures the dynamical effect of the local environment.   In the kaon, the 
$u$-quark is partnered with a significantly heavier partner than in the pion and this shifts the probability to relatively lower $x$ in the kaon.   Our preliminary DSE model 
calculation~\cite{Nguyen:inprep10,Nguyen_PhD09,Holt:2010vj} is shown in Fig.~\ref{fig:pi_DSE+ratio}  along with available Drell Yan  data~\cite{Badier:1980jq,Badier:1983mj}.    Here we include only the leading two invariants of the pion BS amplitude, $E(q, P)$ and $F(q, P)$, where $q$ is $q \bar q$ relative momentum.   For both amplitudes only the lowest Chebychev moment in $q\cdot P$ is employed .   This variable does not occur  in static quantum mechanics, nor in the Nambu--Jona-Lasinio point-coupling field theory model~\cite{Shigetani:1993dx} which also neglects the $q^2$ dependence.   We do not make such a point meson approximation here;  the $q^2$ dependence comes from the BSE solutions.      Nevertheless, the essential features of the ratio $u_K/u_\pi$  are adequately reproduced by a generalized Nambu--Jona-Lasinio model~\cite{Holt:2010vj}.


\begin{theacknowledgments}
The authors would like to thank  C. D. Roberts, S. J. Brodsky and I. Cloet for helpful conversations
and suggestions.   PCT thanks the staff of the CSSM of the University of Adelaide for their warm hospitality, and especially A. W. Thomas for a long acquaintance and the many years of achievement that precipitated this Symposium.  This work has been partially supported by  the U.S. National Science Foundation under grant no. \ PHY-0903991.
\end{theacknowledgments}



\bibliographystyle{aipproc}   

\bibliography{refsPM,refsPCT,refsCDR,refs,refsMAP}

\IfFileExists{\jobname.bbl}{}
 {\typeout{}
  \typeout{******************************************}
  \typeout{** Please run "bibtex \jobname" to optain}
  \typeout{** the bibliography and then re-run LaTeX}
  \typeout{** twice to fix the references!}
  \typeout{******************************************}
  \typeout{}
 }

\end{document}